\documentclass{rrparticle}

\newtheorem{thm}{Theorem}[section]
\newtheorem{prop}[thm]{Proposition}

\newtheorem{rem}[thm]{Remark}

\newtheorem{eg}[thm]{Example}

\numberwithin{equation}{section}

\def\a{\alpha}
\def\b{\beta}

\def\l{\lambda}

\def\n{\,\vert\,}

\def\ca{{\mathcal{A}}}
\def\cb{{\mathcal{B}}}

\def\cg{{\mathcal{G}}}

\def\cj{{\mathcal{J}}}
\def\ck{{\mathcal{K}}}
\def\cl{{\mathcal{L}}}
\def\cm{{\mathcal{M}}}

\def\cp{{\mathcal{P}}}

\def\cu{{\mathcal{U}}}

\def\li{\langle}
\def\ri{\rangle}
\def\n{\ \vert\ }

\def\ms{\medskip}

\def\ni{\noindent}
\def\ti{\tilde}
\def\p{\partial}

\def\diag{{\rm diag}}

\def\C{\mathbb{C}}

\def\N{\mathbb{N}}

\def\R{\mathbb{R} }
\def\Z{\mathbb{Z}}

\def\rd{{\rm d\/}}

\newcommand{\bpm}{\begin{pmatrix}}
\newcommand{\epm}{\end{pmatrix}}
\newcommand{\bca}{\begin{cases}}
\newcommand{\eca}{\end{cases}}

\title{New Hierarchies of Derivative nonlinear Schr\"{o}dinger-Type Equation}
\author[${}$]{Zhiwei Wu}
\author[${}$*]{Jingsong He}
\affil[${}$]{Department of Mathematics, Ningbo University, Ningbo, Zhejiang, 315211, P.\ R.\ China \\$^*${Corresponding author, Email:
\em hejingsong@nbu.edu.cn, jshe@ustc.edu.cn}}

\keywords{Lie algebra splitting, DNLS-type equations, nonlocal reduction}

\begin{document}
\maketitle

\begin{abstract}
{\it Abstract.} We generate hierarchies of derivative nonlinear Schr\"odinger-type equations and their nonlocal extensions from Lie algebra splittings and automorphisms. This provides an algebraic explanation of some known reductions and newly established nonlocal reductions in integrable systems.
\end{abstract}

\section{Introduction: the Derivative NLS-type equations}

Nonlinear Schr\"{o}dinger (NLS) equation is one of the most important examples in  soliton theory and its applications. The focusing NLS equation is of the form:
\begin{equation}\label{nls}
q_t=\frac{i}{2}(q_{xx} +2 |q|^2q).
\end{equation}

NLS equation is a common model of  the propagation of short pulses in nonlinear optical fibers and waves in deep water \cite{AC91}. In nonlinear optics, the linear term in the right hand side of \eqref{nls} denotes the second order dispersion, whereas the nonlinear term accounts for the self-phase-modulation effect. The generic NLS equation \eqref{nls} can be modified to other nonlinear partial differential equations having mathematical and physical meaning; see, for example, a series of papers reviewing the aplication of NLS-type equations to the study of localized optical structures \cite{Malomed2005,Kartashov2011,Chen2012,Grelu2012,Mihalache2012,Leblond2013,Frantzeskakis2014,Malomed2014,Mihalache2015} and Bose-Einstein condensates \cite{Giorgini2008,Frantzeskakis2010,Radha2014,Nicolin2014,Mihalache2014,Radha2015,Nicolin2015,Bagnato2015}.  One of these nonlinear evolution equations, which is of much interest in physical settings, is the so-called derivative nonlinear Schr\"{o}dinger (DNLS) equation  \cite{F84}. By considering different spectral problems (or the $x$-part of Lax pairs), there exist three types of DNLS equations, including the first type (or DNLSI equation) \cite{KN78}:
\begin{equation}\label{ds1}
q_t=\frac{1}{2}(q_{xx}i-(|q|^2q)_x),
\end{equation} 
the second type (or DNLSII equation) \cite{CLL79}:
\begin{equation}\label{ds2}
q_t=\frac{1}{2}(q_{xx}i-|q|^2q_x),
\end{equation}
and the third type (or DNLSIII equation) \cite{GI83}:
\begin{equation}\label{ds3}
q_t=\frac{i}{2}q_{xx}+\frac{1}{2}q^2\bar{q}_x+\frac{i}{4}|q|^4q.
\end{equation}

These equations are used as generic models in plasma physics \cite{M76, R71},
in optical waveguides with the self-steepening effect \cite{AL83, TJ81} and in fluid mechanics \cite{J77}.  In particular, the DNLSII equation has been verified in optical experiments in 2007 \cite{MMW07}, proving that the nonlinear term $|q|^2q_x$ in the DNLSII equation, as well as the term $(|q|^2q)_x$ in the DNLSI equation,  adequately describe the self-steepening effect. The last term in the DNLSIII equation represents the quintic (non-Kerr) nonlinear effect in optical fibers. It is well known that there exist gauge transformations, which involve tedious integrations, between the NLS and DNLS equations, see Refs. \cite{KSS95, K84, WS83}. Besides the very similar but different spectral problems for these four nonlinear evolution equations, they have a new key feature,  namely they possess the so-called rogue wave solutions. This has been shown in a series of works by using the Darboux transformation method; see, for instance, Refs. \cite{ AEK85, GZ14, HZW13, P83, XH12, XHW11, ZG14, ZGX14, ChenMihalache2015}. Recently, the rogue wave solutions of the NLS equation have been observed in optical fibers and water tanks \cite{CHA11, HGZC14, KF10}.

These similarities and  common physical relevance between NLS and DNLS equations pose a natural question:  Is there an unified scheme to generate these equations? The answer of this question will help us a deeper understanding of these nonlinear evolution equations and their integrability properties, such as symmetries of the solution spaces, conserved quantities, and Hamiltonian formulation.

In this paper, we will give a systematic method to generate DNLS-type equations and explain the relations in terms of algebra structure from Lie algebra splitting theory (see details of this theory in the next section). In fact, we get hierarchies of DNLS-type equations that generalize the known results (DNLSI, DNLSII, and DNLSIII equations). These results show the compatibility of reductions of higher flows in the generalized Kaup-Newell (KN) hierarchies (see Theorem \ref{gkn}). From this point of view, there is a natural way to study the symmetries of the solution spaces for these equations-- the B\"{a}cklund transformation. Moreover, higher dimension generation at the Lie algebra level will induce the multi-component soliton equations such as vector 
DNLS-type equations. Finding the Lax pair of given equations (if they are integrable) is always a difficult problem in integrable systems. In order to get NLS and DNLS-type hierarchies from Lie algebra splittings, a highly non-trivial  step is to construct suitable involutions (or automorphisms), although this theory is well established in the literature.

Recently, Ablowitz and Musslimani introduced a new type of reduction ($r(x)=-\bar{q}(-x)$) in the Ablowitz-Kaup-Newell-Segur(AKNS) system, and obtained a nonlocal nonlinear Schr\"{o}dinger equation (nonlocal NLS equation)  \cite{AM13}. This reduction is nonlocal because the value of $r$ at $x$ is related to the value of $\bar{q}$ at $-x$. Such equation is $\mathcal{PT}$-symmetric, and has properties of classical soliton equation, for example, it admits a Lax pair and can be solved by inverse scattering method. In a recent paper \cite{V14}, it was studied the algebraic structure and Hamiltonian formulation of the nonlocal NLS equation. The basic idea to introduce this reduction is to construct newly-established $\mathcal{PT}$-symmetric potentials in non-Hermitian quantum mechanics \cite{BB98}, such that these non-Hermitian operators have real-valued spectra. Recently, besides the theoretical aspects \cite{CGM80, DDT, DDT01, H92}, there have been experimental evidences of $\mathcal{PT}$-symmetry in optical waveguides \cite{G09, RM10}, photonics lattices \cite{RB12,RM13}, plasmonics \cite{B11}, optical metamaterials \cite{FX13},  microcavities \cite{BD12, pengnp2014},  microresonators \cite{changnp2014}, etc.

Basing on the importance of $\mathcal{PT}$-symmetric potentials in physics, it is natural to ask whether we can add this nonlocal reduction in hierarchies of DNLS-type equations in an unified way. We solve this problem from Lie algebra splitting point of view. We need to point out here that in the nonlocal reduction case, the standard splitting theory cannot be applied directly. But we can still construct recursive formula from a special automorphism on the loop group to generate Lax pairs for hierarchies of nonlocal DNLS-type equations. Although the construction is totally algebraic, we believe that these new nonlocal equations would have a good physical relevance basing on the known results on DNLS and nonlocal NLS equations. Moreover, the new involution and automorphism give an algebraic explanation of above nonlocal reduction.

This paper is organized as follows. In Sec. 2, we give a brief review of Lie algebra splitting theory and how to use this scheme to derive defocusing and focusing NLS equations. Then in Sec. 3, we derive generalized KN hierarchies and three hierarchies of DNLS-type equations from Lie algebra splittings, hence we give an algebraic explanation of the relations among DNLS-type equations. In Sec. 4, ``defocusing" DNLS-type equations are studied. We focus in Sec. 5 on nonlocal reductions and show that even flows in the generalized KN hierarchies derived in Sec. 2 admit nonlocal reductions ($r(x, t)=\pm i\bar{q}(-x, t)$), which give rise to the nonlocal DNLS-type equations. We show that the nonlocal constrains come from a special automorphism on the background algebra. New integrable hierarchies are derived from pure algebraic generalization. We discuss in the last section the obtained results and some future research in this area.

\section{Lie algebra splitting}\label{sp}

Integrable systems can be derived from Lie algebra splitting and there have been a series of works in the literature, see, for instance, Refs. \cite{AKNS, DS84, S84, SW85, TU11}. In this section, we give a quick overview of the general scheme of Lie algebra splitting theory. Then we will use it to construct hierarchies of DNLS type in the following sections.

Let $L$ be a compact Lie group, and $\cl$ its Lie algebra. There exist two subgroups $L_+$ and $L_-$  of $L$ such that $L_+\cap L_-= \{e\}$, where $e$ is the identity element in $L$. And in the Lie algebra level, $\cl=\cl_+\oplus \cl_-$ as a direct sum of linear spaces. Then $(\cl_+,\cl_-)$ is called a \emph{splitting} of $\cl$. A \emph{vacuum sequence} $\cj=\{J_1, J_2, \cdots\}$ is a sequence of commuting elements in $\cl_+$, where $J_i$ is an analytic function of $J_1$ in the enveloping algebra of $\cl$. Let $\pi_+$ be the projection of $\cl$ onto $\cl_+$ with respect to the decomposition $\cl=\cl_+\oplus \cl_-$. Then the phase space is of the form:
\begin{equation}\label{f}
\cm=\pi_+(g_-J_1g_-^{-1}), \quad g_-\in L_-.
\end{equation}
\begin{thm} (\cite{TU11}) \label{sa} Given $\xi: \R \rightarrow \cm$, there exists a unique $Q_j(\xi) \in \cl$ such that:
\begin{equation}\label{q}
\begin{cases}
[\p_x+\xi, Q_j(\xi)]=0, \\
Q_j(J_1)=J_j, \quad Q_j(\xi)=M_jJ_jM_j^{-1}, \quad M_j \in L_-.
\end{cases}
\end{equation}
\end{thm}

The \emph{$j$-th flow in the $L$-hierarchy} generated by the splitting $(\cl_+, \cl_-)$ and vacuum sequence $\cj$ is the following evolution of $\xi$:
$$
\xi_{t_j}=[\p_x+\xi, (Q_j(\xi))_+].
$$

There is a natural $L_-$ action on the space of solutions.  Let $L(GL(n, \C))$ be the group of smooth loops from $S^1$ to $GL(n, \C)$, if $L$ is a subgroup of $L(GL(n, \C))$, then the rational elements in $L_-$ can be computed explicitly, moreover, elements with one or two poles often give rise to B\"{a}cklund transformation \cite{TU00}. Hence the loop group splitting presents a natural way to construct new solutions from a given solution. Moreover, the construction is algebraic, therefore, we do not need to solve ordinary equations, which is usually the case in Darboux transformation method.

Next we discuss soliton equations constructed from Lie group $G$ and involution $\sigma$. Denote $L(G)$ to be the group of smooth loops from $S^1$ to $G$, let $\cl(\cg)$ be its Lie algebra. Elements in $\cl(\cg)$ can be written as a power series of $\l$: $$A(\l)=\sum_{i}A_i \l ^i, \quad A_i \in \cg.$$  Let $\sigma$ be an involution of $G$, $\rd_e \sigma$ induces an involution on $\cg$ which is complex linear, where $e$ is the identity element in $G$. Without ambiguity, we still use $\sigma$ to denote the involution on $\cg$ henceforth. Let $\ck$ and $\cp$ be the eigenspace of $\sigma$ on $\cg$ of eigenvalue $1$ and $-1$, respectively. Then
$$[\ck, \ck] \subset \ck, \quad [\ck, \cp] \subset \cp, \quad [\cp, \cp] \subset \ck.$$
Let
$$\cl_\sigma(\cg)=\{A(\l) \in \cl(G) \n \sigma(A(-\l))=A(\l)\}.$$
Then $A(\l)=\sum_{i}A_i \l^i \in \cl_\sigma(\cg)$  if and only if
$$
\begin{cases}
A_i \in \ck, \quad i \quad \text{even}, \\
A_i \in \cp, \quad i  \quad \text{odd}.
\end{cases}
$$
Let $\cl_\sigma(\cg)_+$ and $\cl_\sigma(\cg)_-$ be two subalgebras of $\cl_\sigma(\cg)$ such that $$\cl_\sigma(\cg)=\cl_\sigma(\cg)_+\oplus \cl_\sigma(\cg)_-$$ as a direct sum of linear spaces. Then $(\cl_\sigma(\cg)_+, \cl_\sigma(\cg)_-)$ is a splitting of $\cl_\sigma(\cg)$. Choosing a vacuum sequence $\cj=\{J_1, J_2, \cdots, \} \in \cl_{\sigma}(\cg)_+$, the phase space for the evolution equations is defined as follows:
\begin{equation}\label{fa}
\cm=\pi_+(g_-J_1g_-^{-1}), \quad g_- \in L_\sigma(\cg)_-,
\end{equation}
where $\pi_+$ is the projection of $\cl_\sigma(\cg)$ onto $\cl_\sigma(\cg)_+$ with respect to the splitting $\cl_\sigma(\cg)=\cl_\sigma(\cg)_+\oplus\cl_\sigma(\cg)_-$.

\begin{thm} (\cite{TU11}) \label{sc} Given $\xi \in C^{\infty}(\R, \cm)$, there exists a unique $Q_j(\xi) \in \cl_\sigma(\cg)$  for any $j \geq 1$ such that
\begin{equation}\label{qa}
\begin{cases}
[\p_x+\xi, Q_j(\xi)]=0, \\
Q_j(J_1)=J_j \\
Q_j(\xi) \  \text{is conjugate to} \ J_j.
\end{cases}
\end{equation}
\end{thm}
The \emph{$j$-th flow in the $(G, \sigma)$-hierarchy} is
\begin{equation}\label{jf}
[\p_x+\xi, \p_{t_j}+(Q_j(\xi))_+]=0.
\end{equation}

\begin{eg} \label{efn} {\bf [$SU(2)$-hierarchy and focusing NLS equation]} \hfil \par

Let
$$
\cl(su(2))=\left\{A(\l)=\sum_iA_i\l^i \mid A_i \in su(2), A(\l)=-\overline{A(\bar{\l})}^t\right\},
$$
and
$$
\begin{cases}
\cl_+(su(2))=\{\sum_{i \geq 0}A_i\l^i \mid A_i \in su(2)\}, \\
\cl_-(su(2))=\{\sum_{i <0} A_i \l^i \mid A_i \in su(2)\}.
\end{cases}
$$
Then $(\cl_+(su(2)),\cl_-(su(2)))$ is a splitting of $\cl(su(2))$. Let $a=\diag(i. -i)$, and $J_1=a\l$, then $\cj=\{a\l^i \mid i \geq 1\}$ is a vacuum sequence. The phase space defined by \eqref{f} is of the form:
$$
\xi=J_1+u=a\l+\bpm 0 & q \\ -\bar{q} & 0\epm, \quad q \in C^{\infty}(\R, \C).
$$
We solve $Q(u, \l)=a\l+Q_0+Q_{-1}\l^{-1}+\cdots \in \cl(su(2))$ from \eqref{q}, and get
\begin{align*}
& Q_0=u, \quad Q_{-1}=\frac{i}{2}\bpm -|q|^2 & q_x \\ \bar{q}_x & |q|^2 \epm, \\
& Q_{-2}=\frac{1}{4}\bpm q_x\bar{q}-q\bar{q}_x & -q_{xx}-2|q|^2q \\ \bar{q}_{xx}+2|q|^2\bar{q} & q\bar{q}_x-q_x\bar{q}\epm.
\end{align*}
The second flow $u_t=[\p_x+u, Q_{-1}]=[Q_{-2}, a]$ is the focusing NLS equation \eqref{nls}.
\end{eg}

\begin{eg} \label{en} {\bf [$U(1, 1)$-hierarchy and defocusing NLS equation]}\hfil \par

Let $U(1, 1)$ be the subgroup of $SL(2, \C)$ preserving the bilinear form in $\C^2$:
$$\li X, Y \ri=\bar{X}^t I_{1, 1} Y, \quad X, Y \in \C^2.$$
Let $u(1, 1)$ be the Lie algebra for $U(1, 1)$. Then
$$u(1, 1)=\{g \in sl(2, \C) \mid \bar{g}I_{1, 1}+I_{1, 1}g=0\}=\left\{\bpm \a i & \b \\ \bar{\b} & -\a i \epm \mid \a \in \R, \b \in \C \right\}.$$
Consider the splitting $(\cl_+(u(1, 1)), \cl_-(u(1, 1)))$ of $\cl(u(1, 1))$ such that
$$
\begin{cases}
\cl_+(u(1, 1))=\{\sum_{i \geq 1}A_i\l^i \mid A_i \in u(1, 1)\}, \\
\cl_-(u(1, 1))=\{\sum_{i < 0}A_i\l^i \mid A_i \in u(1, 1)\}.
\end{cases}
$$
Choosing the vacuum sequence $\cj=\{a\l^i \mid i \geq 1\}$, where $a=\diag(i, -i)$, the phase space is of the form:
$$
a\l+u=a\l+\bpm 0 & q \\ \bar{q} & 0 \epm, \quad q \in C^{\infty}(\R, \C).
$$
The second flow is the defocusing NLS equation:
\begin{equation}\label{fnls}
q_t=\frac{i}{2}(q_{xx}-2|q|^2q).
\end{equation}
\end{eg}

\section{Hierarchies of DNLS-type equations}

The original construction of DNLSI, II, and III equations is based on different constraints on the Kaup-Newell (KN) system \cite{KN78}. From the principle grading of affine Kac-Moody algebra of type $\hat{A}_1(sl_2)$, the algebraic dressing method can be used to construct soliton solutions for the KN system \cite{G02}. The standard method to solve the DNLSI, II, and III equations is the {\it Darboux transformation}. For each case, we need to choose a special form of Darboux transformation compatible with the corresponding reduction to generate new solutions from a given solution. Sometimes this process would be rather difficult if we do not have enough information from the Lax pair.

In this section we will systematically derive DNLS-type equations from splitting of Lie algebras. In this manner, we would not need to choose special forms of Darboux transformation case by case in order to satisfy the constraint every time we see a 
DNLS-type equation. Instead, the Lie algebra splitting method will provide a standard process to construct the Darboux transformation \cite{TU00}.

To get a through story, we also start from the KN system and use the Lie algebra splitting to generate generalized KN hierarchies. Later in this section, we will show how to derive the DNLS-type equations directly from different group actions.
\subsection{The Kaup-Newell system}
Let $G=SL(2, \C)$, and $L(SL(2, \C))$ be the group of smooth loops from $S^1$ to $SL(2, \C)$, and $\cl(sl(2, \C))$ its Lie algebra. Consider the following splitting of $\cl(sl(2, \C))$:
\begin{align*}
& \cl_+(sl(2, \C))=\{\sum_{i \geq 1}A_i \l^i \n A_i \in sl(2, \C)\}, \\
& \cl_-(sl(2, \C))=\{\sum_{i \leq 0} A_i \l^i \n A_i \in sl(2, \C)\}.
\end{align*}
Define an involution $\sigma$ on $sl(2, \C)$ as following:
\beq\label{s}
\sigma (A)=I_{1, 1}AI_{1, 1}^{-1}, \quad I_{1,1}=\diag(1, -1).
\eeq
Let $\ck$ and $\cp$ be the $1$ and $-1$ eigenspace of $\sigma$, respectively. Then
$$\ck=\diag(\a, -\a), \quad \cp=\bpm 0 & \beta \\ \eta & 0 \epm, \quad \a, \beta, \eta \in \C. $$
Furthermore, $\sigma$ induces an involution on $\cl(sl(2, \C))$ such that
$$\sigma (A(\l))=I_{1,1}A(-\l)I_{1, 1}^{-1}.$$
Let $\cl=\cl_{\sigma}(sl(2, \C))$ be the subalgebra of $\cl(sl(2, \C))$ consisting of fixed points of $\sigma$, and
$$ \cl_+=\cl \cap  \cl_+(sl(2, \C)), \quad \cl_-=\cl \cap  \cl_-(sl(2, \C)). $$
Let $a=\diag(i, -i)$, and vacuum sequence $\cj=\{J_1, J_2\cdots\}$ such that $J_i=a\l^{2j}$. Then the phase space is of the form: $a\l^2+\cp\l$.

Given $u=\bpm 0 & q \\ r & 0\epm \in \cp$, there exists a unique $Q(u, \l)=a\l^2+Q_1\l+Q_0+Q_{-1}\l^{-1}+\cdots$, where $Q_i \in sl(2, \C)$, such that
$$
\bca
[\p_x+a\l^2 +u \l, Q(\l)]=0, \\
Q(\l)^2=-\l^4.
\eca
$$
The $j$-th flow \eqref{jf} in the \emph{$(SL(2, \C), \sigma)$-hierarchy} is
\beq\label{KN1}
u_{t_j}=[\p+a\l^2+u\l, (Q(\l)\l^{2(j-1)})_+]=(Q_{3-2j})_x.
\eeq
From a direct computation, we have
\begin{align*}
& Q_1=u, \quad Q_0=\frac{i}{2}\diag(qr, -qr), \quad Q_{-1}=\frac{1}{2}\bpm  0 & q_x i+q^2r \\ -r_x i + qr^2 \epm, \\
& Q_{-2}=\frac{1}{8}\bpm 2(qr_x-q_xr)+3q^2r^2i & 0 \\ 0 & 2(q_xr-qr_x)-3q^2r^2i \epm,\\
& Q_{-3}=-\frac{1}{8}\bpm 0 & 2q_{xx}-6qrq_xi-3q^3r^2 \\ 2r_{xx}+qrr_x i-3q^2r^3 & 0\epm.
\end{align*}
The second flow is the KN system \cite{KN78},
$$
\bca
q_t=\frac{1}{2}(q_{xx}i+(q^2r)_x), \\
r_t=\frac{1}{2}(-r_{xx}i+(qr^2)_x).
\eca
$$
The third flow is:
$$
\bca
q_t=-\frac{1}{4}q_{xxx}+\frac{3i}{4}(qrq_x)_x+\frac{3}{8}(q^3r^2)_x, \\
r_t=-\frac{1}{4}r_{xxx}-\frac{3i}{4}(qrr_x)_x+\frac{3}{8}(q^2r^3)_x.
\eca
$$

Let $G=SU(2)$ instead of $SL(2, \C)$, following the same algorithm, we get the phase space of the form $a \l^2 +\bpm 0 & q \\ -\bar{q} & 0\epm \l$, i.e., $r=-\bar{q}$. Then the second flow in the $(SU(2), \sigma)$-hierarchy is the DNLSI equation \eqref{ds1}.
\subsection{Derivative nonlinear Schr\"{o}dinger equation III}
Different splittings of $\cl_\sigma(sl(2, \C))$ will lead to new integrable hierarchies different from the KN hierarchy:
$$
\bca
\cl_\sigma(sl(2, \C))_+=\{\sum_{i \geq 0}A_i \l^i \in \cl_\sigma(sl(2, \C)) \}, \\
\cl_\sigma(sl(2, \C))_-=\{\sum_{i \leq -1}A_i \l^i \in \cl_\sigma(sl(2, \C)) \}.
\eca
$$
Let $J_1=a\l^2 \in \cl_\sigma(sl(2, \C))_+$, by \eqref{fa}, the phase space is of the form:
$$(g_-J_1g_-^{-1})_+=a\l^2+\bpm 0 & q \\ r & 0 \epm\l+\frac{i}{2}\bpm qr & 0 \\ 0  & -qr \epm, \quad q, r \in \C. $$
Given $u=\bpm 0 & q \\ r & 0 \epm$, let $P_0=\frac{i}{2}\bpm qr & 0 \\ 0  & -qr \epm$, by Theorem \ref{sc}, there exists a unique $Q(u, \l)=a\l^2+Q_1\l+\cdots \in \cl_\sigma(sl(2, \C))$ such that
$$
\bca
[\p_x+a\l^2+u\l+P_0, Q(\l)]=0, \\
Q(\l)^2=-\l^4.
\eca
$$
The $j$-th flow in this hierarchy is
$$
[\p_x+a\l^2+u\l+P_0, \p_{t_j}+(Q(\l)\l^{2(j-1)})_+]=0,
$$
which is equivalent to
\beq\label{jc}
u_{t_j}=(Q_{3-2j})_x+[P_0, Q_{3-2j}]+[u, Q_{2-2j}].
\eeq
We call these equations belong to the \emph{$(SL(2, \C), \sigma)$-hierarchy of type II}.

From a direct computation,
\begin{align*}
& Q_1=u, \quad Q_0=P_0, \quad Q_{-1}=\frac{i}{2}\bpm 0 & q_x \\ -r_x & 0 \epm, \\
& Q_{-2}=\frac{1}{8}\bpm 2(qr_x-q_xr)-q^2r^2i & 0 \\ 0 &  2(q_xr-qr_x)+q^2r^2i\epm, \\
& Q_{-3}=\frac{1}{8}\bpm 0 & -2q_{xx}-2iq^2r_x-q^3r^2 \\ 2r_{xx}+2iq_xr^2-q^2r^3 \epm, \\
& Q_{-4}=\frac{i}{8}\bpm -qr_{xx}-q_{xx}r+q_xr_x-\frac{1}{2}q^3r^3 & 0 \\ 0 & qr_{xx}+q_{xx}r-q_xr_x+\frac{1}{2}q^3r^3\epm
\end{align*}
The second flow is the following coupled system (the Gerdjikov-Ivanov system \cite{GI83}):
\beq\label{gi1}
\bca
q_t=\frac{i}{2}q_{xx}-\frac{1}{2}q^2r_x+\frac{i}{4}q^3r^2, \\
r_t=-\frac{i}{2}r_{xx}-\frac{1}{2}q_xr^2-\frac{i}{4}q^2r^3.
\eca
\eeq
The 3rd flow is a coupled third-order system of partial-differential equations (PDEs):
$$
\bca
q_t=-\frac{1}{4}q_{xxx}-\frac{3i}{4}qq_xr_x-\frac{3}{8}q^2r^2q_x, \\
r_t=-\frac{1}{4}r_{xxx}+\frac{3i}{4}rq_xr_x-\frac{3}{8}q^2r^2r_x.
\eca
$$

The system \eqref{gi1} admits the constant $r=-\bar{q}$, hence becomes the Gerdjikov-Ivanov (or DNLSIII) equation:
\beq\label{gi2}
q_t=\frac{i}{2}q_{xx}+\frac{1}{2}q^2\bar{q}_x+\frac{i}{4}|q|^4q.
\eeq

\begin{prop} Equations belonging to the $(SL(2, \C), \sigma)$-hierarchy of type II admits the constraint $r=-\bar{q}$.
\end{prop}

We prove this from the algebraic structure. Let $\tau$ be a group automorphism of $SL(2, \C)$ such that $\tau(A)=\bar{A}^{-1}$. Let $U$ be the space of fixed points of $\tau$, and $\cu$ its Lie algebra. Then $\cu$ is a \emph{real form} of $sl(2, \C)$. Let $\cl_{\tau, \sigma}(sl(2, \C))$ be the subalgebra of  $\cl_{\sigma}(sl(2, \C))$ such that,
$$\cl_{\tau, \sigma}(sl(2, \C))=\cl_\sigma(su(2))=\{A(\l) \in \cl_{\sigma}(sl(2, \C)) \mid \tau(A(\bar{\l}))=A(\l)\}. $$
In other words, $A(\l)=\sum_{i}A_i\l^i \in \cl_\sigma(su(2))$ if and only if
$$ A_i=
\bca
\bpm  r i & 0 \\ 0 & -r i  \epm, &\quad  r \in \R, \quad i \quad \text{even}, \\
\bpm 0 & q \\ -\bar{q} & 0\epm, & \quad q \in \C, \quad i  \quad \text{odd}.
\eca
$$

Consider the splitting $(\cl_+, \cl_-)$ of $\cl_\sigma(su(2))$ such that
$$\cl_+=\{\sum_{i \geq 0} A_i \l^i  \in \cl_\sigma(su(2))\}, \cl_-=\{\sum_{i \leq -1} A_i \l^i  \in \cl_\sigma(su(2))\}. $$
Choosing the same vacuum sequence as in the $(SL(2, \C), \sigma)$-hierarchy of type II,  then the $j$-th flow in the $(SU(2), \sigma)$-hierarchy of type II is the constraint case of \eqref{jc} with the constraint $r=-\bar{q}$. This prove the Proposition.

In particular, the second flow in the $(SU(2), \sigma)$-hierarchy of type II is the DNLSIII equation \eqref{ds3}.  And the third flow is
$$
q_t=-\frac{1}{4}q_{xxx}+\frac{3}{4}iq|q_x|^2-\frac{3}{8}|q|^4q_x.
$$
\begin{rem}
The reason we study the DNLSIII equation before the DNLSII is because the splitting of these two hierarchies are highly related and just different by a shift of free term in $\l$ on the loop algebra. On the other hand, the DNLSII is a more general case whose algebra structure is much more complicated.
\end{rem}
\subsection{Derivative nonlinear Schr\"{o}dinger equation II}
Let $\ca$ be the subalgebra of diagonal matrices in $sl(2, \C)$, and $\cb$ be a linear map on $\ca$. Consider the following splitting $(\cl_+, \cl_-)$ of  $\cl_\sigma(sl(2, \C))$ such that for $\sum_{i}A_i\l^i \in \cl_\sigma(sl(2, \C))$:
$$(\sum_{i}A_i\l^i)_{\cl_+}=\sum_{i \geq 1}A_i\l^i+A_0-\cb(A_0).$$
We use the same vacuum sequence $\cj=\{a\l^2, a\l^4, \cdots \}$ as in the previous cases and let
$$\cb(A)=(1-2\a i)A, \quad \a \in \C, \quad A \in \ca.$$

By definition and direct computation, the phase space is of the form:
$$\cm=a\l^2+u\l+P_0=a\l^2+\bpm 0 & q \\ r & 0\epm \l-\a\bpm qr & 0 \\ 0& -qr \epm.$$
Let $Q(u, \l)=a\l^2+Q_1\l+Q_0+\cdots$ be the unique element in $\cl_\sigma(sl(2, \C))$, such that:
$$
\bca
[\p_x+a\l^2+u\l+P_0, Q(u, \l)]=0, \\
Q(u, \l)^2=-\l^4.
\eca
$$
Then the $j$-th flow in the \emph{$(SL(2, \C), \sigma)$-hierarchy of twisted by $\cb$} is
$$
[\p_x+a\l^2+u\l+P_0, \p_{t_j}+(Q(u, \l)\l^{2(j-1)})_+]=0.
$$
Or equivalently,
\beq\label{KN2}
u_{t_j}=(Q_{3-2j})_x+[P_0, Q_{3-2j}]+[u, Q_{2-2j}-\cb(Q_{2-2j})].
\eeq
We give the first several terms of $Q(u, \l)$:
\begin{align*}
& Q_1=u, \quad Q_0=\frac{i}{2}\bpm qr & 0 \\ 0 & -qr \epm, \\
&Q_{-1}=\frac{1}{2}\bpm 0 & q_x i- (2i\a-1)q^2 r \\ -r_x i  - (2i \a -1)q r^2\epm, \\
& Q_{-2}=\frac{1}{4}\diag(qr_x-q_x r+(4\a+\frac{3}{2}i)q^2r^2, q_xr-qr_x-(4\a+\frac{3}{2}i)q^2r^2), \\
& Q_{-3}=\bpm 0 & b  \\ c & 0 \epm,  \quad Q_{-4}=\bpm d & 0 \\ 0 & -d \epm,
\end{align*}
where
\begin{align*}
& b=-\frac{1}{4}q_{xx}+\frac{\a}{2} q^2 r_x+(\frac{3\a}{2}+\frac{3 i}{4})qrq_x+(\frac{3}{8}-\frac{3 \a i}{2}-\a^2)q^3r^2,  \\
&c=-\frac{1}{4}r_{xx}-\frac{\a}{2}r^2q_x+(\frac{3\a}{2}+\frac{3 i}{4})qrq_x-(\a^2+\frac{3 \a i}{2}-\frac{3}{8})q^2r^3, \\
& d=-\frac{i}{8}(qr_{xx}+rq_{xx}-q_xr_x)+\frac{3}{8}(2 \a i-1)(qr^2q_x-q^2rr_x) \\
& \quad -\frac{3}{2}(\a^2 i - \a -\frac{5 i}{24})q^3r^3.
\end{align*}
The second flow is:
\beq\label{aa}
\bca
q_t=\frac{i}{2}q_{xx}-(i\a-\frac{1}{2})(q^2r)_x-i\a q^2 r_x+(\frac{1}{2}\a-2i\a^2)q^3r^2,\\
r_t=-\frac{1}{2}ir_{xx}-(i\a-\frac{1}{2})(qr^2)_x-i\a q_xr^2+(2i\a^2-\frac{1}{2}\a)q^2r^3.
\eca
\eeq
Similar to the $(SL(2, \C), \sigma)$-hierarchy of type II, if we consider the real form $SU(2)$ of $SL(2, \C)$, then \eqref{aa} becomes the equation:
\beq\label{a}
q_t=\frac{i}{2}q_{xx}+(2\a i-1)|q|^2q_x+(2\a i-\frac{1}{2})q^2\bar{q}_x+(\frac{1}{2}\a-2\a^2 i)|q|^4q,
\eeq
where $\a$ is pure imaginary.

A general form of this equation is introduced in \cite{LP13}. Let $\a=-\frac{i}{4}$. Then \eqref{a} becomes the DNLSII equation \eqref{ds2}:
$$
q_t=\frac{1}{2}iq_{xx}-\frac{1}{2}|q|^2q_x.
$$
\section{Defocusing analogy}

In this section, we study the defocusing DNLS-type equations. It is well known that defocusing NLS \eqref{nls} has only dark soliton solutions, while the focusing NLS \eqref{fnls} have both bright soliton and breather solutions. Both bright and dark soliton solutions of DNLSI have been found \cite{XHW11}. Therefore, it is not quite precise to name ``focusing" and ``defocusing" DNLS equations. But to be consistent with the notation of NLS case as in Examples \ref{efn} and \ref{en}, in this paper, we still call the DNLS-type equation derived from $U(1, 1)$ the \emph{defocusing DNLS}. In this section, we will derive the defocusing DNLSI, II, and III equations. Since it is just following the same scheme we described in the previous sections with rather tedious computations, we skip the process and only list the main results here.

\ms

\ni {\bf The $(U(1, 1), \sigma)$-hierarchy} \

Let $\sigma$ be the involution defined as \eqref{s}, $$\cl_\sigma(u(1, 1)):=\{A(\l) \in \cl(u(1, 1)) \mid \sigma(A(-\l))=A(\l)\},$$
and
$$
\bca
\cl_\sigma(u(1, 1))_+=\{\sum_{i \geq 1}A_i\l^i \in \cl_\sigma(U(1, 1))\}, \\
\cl_\sigma(u(1, 1))_-=\{\sum_{i < 1}A_i \l^i \in \cl_\sigma(U(1, 1))\}.
\eca
$$
Let $\cj=\{a\l^{2j} \mid j \geq 1\}$, where $a=\diag(i, -i)$ be the vacuum sequence. Then the phase space defined by \eqref{fa} is of the form:
$$
a\l^2+u\l=a\l^2+\bpm 0 & q \\ \bar{q} & 0 \epm \l.
$$
Denoting $Q(u, \l)=a\l^2+u\l+Q_0+Q_{-1}\l^{-1}+\cdots \in \cl_\sigma(u(1, 1))$, by \eqref{qa}, we can solve $Q(u, \l)$ uniquely. In particular, the first several terms are:
\begin{align*}
&Q_1=u,  \quad Q_0=\bpm \frac{1}{2}|q|^2 i & 0 \\ 0 & -\frac{1}{2}|q|^2 i \epm \\
&Q_{-1}= \frac{1}{2}\bpm 0 & q_x i+|q|^2 q \\ -\bar{q}_x i+\bar{q}|q|^2 & 0 \epm.
\end{align*}
The second flow is $[\p_x+a\l^2+u\l, a\l^4+u\l^3+Q_0\l^2+Q_{-1}\l]$, which is the defocusing DNLSI equation:
\beq\label{dds}
q_t=\frac{1}{2}(q_{xx}i+(|q|^2q)_x).
\eeq

\ms

\ni {\bf The $(U(1, 1), \sigma)$-hierarchy twisted by $\cb$} \

Let $\cb$ be a linear map on $\ca$, the subalgebra of diagonal matrices in $su(1, 1)$, such that
$$\cb (A) = (1-2\a i)A, \quad Re(\a)=0, \quad A \in \ca.$$
Then the phase space in the $(U(1, 1), \sigma)$-hierarchy twisted by $\cb$ is of the form:
$$
a\l^2+\bpm 0 & q \\ \bar{q} & 0\epm \l-\a\bpm |q|^2 & 0 \\ 0 & -|q|^2\epm.
$$
The second flow is
$$
q_t=\frac{i}{2}q_{xx}+(1-2\a i)|q|^2q_x+(\frac{1}{2}-2\a i)q^2\bar{q}_x+(\frac{1}{2}\a+2|\a|^2 i)|q|^4q.
$$
In particular, when $\a=-\frac{1}{4}i$, we get defocusing DNLSII equation:
\beq\label{dds2}
q_t=\frac{i}{2}q_{xx}+\frac{1}{2}|q|^2q_x.
\eeq

\ms
\ni{\bf The $(U(1, 1), \sigma)$-hierarchy of type II}\

The $(U(1, 1), \sigma)$-hierarchy of type II generated by $a \l^2$ has the phase space of the form:
$$a\l^2+\bpm 0 & q \\ \bar{q} & 0 \epm \l +\frac{i}{2}\bpm |q|^2 & 0 \\ 0  & -|q|^2 \epm.$$
Let $Q(u, \l)=a\l^2+Q_1\l+Q_0+\cdots$, where
\begin{align*}
& Q_1=u, \quad Q_0=\frac{i}{2}\bpm |q|^2 & 0 \\ 0 & -|q|^2 \epm, \quad Q_{-1}=\frac{i}{2}\bpm 0 & q_x \\ -q_x &  0\epm, \\
& Q_{-2}=-\frac{1}{8}\bpm 2(\bar{q}q_x-\bar{q}_xq)+|q|^4 i & 0 \\ 0 & 2(\bar{q}_xq-\bar{q}q_x)-|q|^4 i\epm.
\end{align*}
The second flow is the defocusing DNLSIII equation:
\beq\label{dds3}
q_t=\frac{i}{2}q_{xx}-\frac{1}{2}q^2\bar{q}_x+\frac{i}{4}|q|^4q.
\eeq

In summary, from the construction of these hierarchies, we actually prove the following theorem:
\begin{thm} \label{gkn} The $j$-th flows in generalized KN hierarchies \eqref{KN1}, \eqref{jc}, and \eqref{KN2} admit the constraints $r=\pm\bar{q}$ for each $j \in \N$.
\end{thm}

\section{Nonlocal nonlinear Schr\"{o}dinger equation}

From the discussion of previous sections, we have seen that DNLS-type equations can be derived from Lie algebra splittings and involutions. In this section, we will construct a new type of automorphisms to derive the integrable hierarchies of nonlocal DNLS-type. To be consistent, we start with the algebra structure of nonlocal NLS equation induced in \cite{AM13}, which was derived from a purpose of the construction of $\mathcal{PT}$-symmetrical potentials.

\subsection{Nonlocal NLS equation}
Let $G$ be a compact Lie group and $\cg$ its Lie algebra. Let $\tau$ be an involution on $G$ such that $\tau_{\ast}=\rd_e \tau$ is a conjugate linear involution on $\cg$. To simplify the notation, we still use $\tau$ to denote the involution on $\cg$. Denote $\ti{\tau}$ to be the following involution on $\cl(\cg)$, the loop algebra of $G$:
\beq\label{b}
\ti{\tau}(f(x,\l))=\sum_j\tau(f_j(-x)(-\bar{\l})^{j}), \quad f(x, \l)=\sum_{j}f_j(x)\l^j, f_j(x) \in \cg.
\eeq

Let $J=\bpm 0 & 1\\ 1 & 0 \epm$, define an involution $\tau_1$ on $sl(2, \C)$ as following:
$$\tau_1(A)=-J \bar{A} J^{-1},  \quad A \in sl(2, \C).$$
Let $\ck$ and $\cp$ be the eigenspace of induced involution $\ti{\tau}_1$ of eigenvalue $1$ and $-1$,  respectively. Then
$$
\bca
\ck=\left\{\bpm  a(x) & b(x) \\ c(x) & -a(x)\epm, \quad a(x)=\bar{a}(-x), c(x)=-\bar{b}(-x) \right\}, \\
\cp=\left\{ \bpm   a(x) & b(x) \\ c(x) & -a(x)\epm, \quad a(x)=-\bar{a}(-x), c(x)=\bar{b}(-x) \right\}.
\eca
$$

Let $$\cl_{\ti{\tau}_1}(sl(2, \C))=\{f(x, l) \in \cl(sl(2, \C)) \mid \ti\tau_1(f(x, \l))=f(x, \l)\}.$$
Then $f(x, \l)=\sum_{i}f_i(x)\l^i \in \cl_{\ti{\tau}_1}(sl(2, \C))$ if and only if:
$$
f_i(x) \in
\bca
\cp, \quad  i \  \text{odd}, \\
\ck, \quad  i \  \text{even}.
\eca
$$

Let $a=\diag(i. -i)$, and $u=\bpm 0 & q(x, t) \\ -\bar{q}(-x, t) \epm \in \ck$. From direct computations, we can solve $Q(u, \l)=a\l+\sum_{i=0}^{\infty}Q_{-i}\l^{-i} \in \cl_{\ti{\tau}_1}(sl(2, \C))$ uniquely by following equations:
\beq\label{af}
\bca
[\p_x+a\l+u, Q(u, \l)]=0, \\
Q(u, \l)^2=-I_2.
\eca
\eeq

In particular,
\begin{align*}
& Q_0=u=\bpm 0 & q(x, t) \\ -\bar{q}(-x, t) \epm, \quad Q_{-1}=-\frac{i}{2}\bpm q(x)\bar{q}(-x) & -q_x(x) \\ \bar{q}_x(-x) & -q(x)\bar{q}(-x)\epm, \\
& Q_{-2}=\frac{1}{4}\bpm \bar{q}_x(-x)q(x)+q_x(x)\bar{q}(-x) & -q_{xx}(x)-2q^2(x)\bar{q}(-x) \\ \bar{q}_{xx}(-x)+2q(x)\bar{q}^2(-x) & -\bar{q}_x(-x)q(x)-q_x(x)\bar{q}(-x)\epm.
\end{align*}

The second flow $u_t=[\p_x+a\l+u, a\l^2+u\l+Q_{-1}]=(Q_{-1})_x+[u, Q_{-1}]$ is the nonlocal NLS equation \cite{AM13}:
$$
q_t(x, t)=\frac{i}{2}q_{xx}(x, t)+iq^2(x, t)\bar{q}(-x. t).
$$

Note that the next flow in this hierarchy is the fourth flow instead of the third flow. The reason for this is that $[\p_x+u, Q_{-2j-1}] \in \ck$, while
$[\p_x+u, Q_{-2j}] \in \cp$, which is not compatible with $u_t \in \ck$.

\begin{rem} Although the nonlocal reduction $r(x, t)=-\bar{q}(-x, t)$ can be obtained by choosing special involution $\ti \tau$ such that  the phase space $a\l + u$ is belonging to the fixed points set of $\ti{\tau}$, it can be checked that
$$[\p_x, \ck] \in \cp, \quad [\p_x, \cp] \in \ck.$$
Therefore the standard splitting theory in Sec. \ref{sp} does not work here. But we can still solve the recursive formula similar to \eqref{af} and construct the hierarchy. This is different with all the NLS and DNLS-type equations we dealt with in previous sections of this paper.
\end{rem}

Next we define another involution $\tau_2$ as following:
$$\tau_2(A)=\bar{A}^t, \quad A \in sl(2, \C).$$
Then the second flow generated by $a\l +u$, $u=\bpm 0 & q(x, t) \\ \bar{q}(-x, t) & 0 \epm$ is the other nonlocal NLS equation in \cite{AM13}:
$$
q(x, t)_t=\frac{i}{2}q_{xx}(x, t)-iq^2(x, t)\bar{q}(-x. t).
$$

\subsection{Nonlocal DNLS-type equations}
In the following, we will study nonlocal reductions of the DNLS-type equations. In fact, we can prove the following theorem:

\begin{thm} \label{nn} The even flows in the generalized KN hierarchies \eqref{KN1}, \eqref{jc}, and \eqref{KN2} admits nonlocal  constraints of the type $r(x, t)=\pm i\bar{q}(-x, t)$.
\end{thm}

We prove this theorem by finding the algebra structure for each case then deriving the flows from Lie algebra splitting with certain automorphisms. Since the process for the three types listed above is similar with routine computation, we only discuss the nonlocal DNLSI equation here, and write down the other two types of equations in the end.

First we define an automorphism $\xi$ on $C^{\infty}(\R, sl(2, \C))$ as following:
\beq\label{ae}
\xi(A)(x)=\bpm 0 & i \\ i & 0  \epm \bar{A}(-x) \bpm 0 & i \\ i & 0  \epm,
\eeq
Note that $\xi$ commutes with $\sigma$ defined in \eqref{s}. Moreover, $\xi(A)$ induces an automorphism on $\cl_\sigma(sl(2, \C))$ such that for $f(x, t, \l)=\sum_if_i(x, t)\l^i \in \cl_\sigma(sl(2, \C))$,
\beq\label{ae2}
\ti{\xi}(f(x, t, \l))=\sum_j \xi(f_j(x, t))(i\l)^j.
\eeq

Let $\cg_j$ be the eigenspace of $\xi$ in $C^{\infty}(\R, sl(2, \C))$ with respect to eigenvalue $i^j$, $0 \leq j \leq 3$, and $\cl_{\sigma, \xi}(sl(2, \C))$ the set of fixed points of $\xi(A)$ on $\cl_\sigma(sl(2, \C))$. Note that if $f(x, t, \l)=\sum_j \xi(f_j(x, t))(i\l)^j \in \cl_{\sigma, \xi}(sl(2, \C))$, then
$$f_{4k+j}(x, t) \in \cg_{j}, \quad k \in \Z, \quad 0 \leq j \leq 4. $$

Given $u=\bpm  0 & q(x, t) \\ -i\bar{q}(-x, t)\epm$,  then $a\l^2 +u\l \in \cl_{\sigma, \xi}(sl(2, \C))$.

By direct computation, we can solve $Q(u, \l)=a\l^2+Q_1\l+Q_0+Q_{-1}\l^{-1}+\cdots  \in \cl_{\sigma, \xi}(sl(2, \C))$ uniquely from following recursive formula:
$$
\bca
[\p_x+a\l^2+u\l, Q(u, \l)]=0, \\
Q(u, \l)^2=-\l^4.
\eca
$$
To get the second flow, we need the first several terms:
\begin{align*}
& Q_1=u, \quad  Q_0=\frac{1}{2}\bpm q(x, t)\bar{q}(-x, t) & 0 \\ 0 & -q(x,t)\bar{q}(-x, t) \epm, \\
& Q_{-1}=\frac{1}{2} \bpm 0 & i(q_x(x, t)-q^2(x, t)\bar{q}(-x, t)) \\ \bar{q}_{x}(-x, t)-q(x, t)\bar{q}^2(-x, t) & 0\epm, \\
& Q_{-2}=\diag(w, -w), \\
& w=\frac{i}{8}(2(q(x, t)\bar{q}_x(-x, t)+q_x(x, t)\bar{q}(-x, t))-3q^2(x, t)\bar{q}^2(-x, t)).
\end{align*}
Therefore, the second flow is
$$
u_t =[\p_x+a\l^2+u\l, a\l^4+u\l^3+Q_0\l^2+Q_{-1}\l]=(Q_{-1})_x.
$$
Written in terms of $q$, this equation is one of the nonlocal DNLSI equation:
$$
q_t(x, t)=\frac{i}{2}q_{xx}(x, t)-2 i q(x, t)\bar{q}(-x, t)q_x(x, t)+\frac{i}{2}q^2(x, t)\bar{q}_x(-x, t).
$$

The next flow in this nonlocal DNLSI hierarchy is the fourth flow. The reason is because that $[\p_x, \cg_3] \in \cg_1$, hence $[\p_x, Q_{-(4k-3)}]$ generates a flow for $u \in \cg_1$, for each $k \in \Z$. And this is the $2k$-th flow in the hierarchy.

To get the nonlocal reduction $r(x, t)=i\bar{q}(-x, t)$, we need to consider the following automorphism on $C^{\infty}(\R, sl(2, \C))$:
\beq\label{ae3}
\xi(A)(x)=\bpm 0 & i \\ -i & 0  \epm \bar{A}(-x) \bpm 0 & -i \\ i & 0  \epm,
\eeq
and the remaining process is similar.

To summary, with the help of splitting theory, we are able to get the following equations:
\begin{enumerate}
\item The nonlocal DNLSI equation:
\beq\label{nds1}
q_t(x, t)=\frac{i}{2}q_{xx}(x, t)\pm2 i q(x, t)\bar{q}(-x, t)q_x(x, t)\mp\frac{i}{2}q^2(x, t)\bar{q}_x(-x, t).
\eeq
\item The nonlocal DNLSII equation:
\beq\label{nds2}
q_t(x, t)=\frac{i}{2}q_{xx}(x, t)\pm\frac{i}{2}q(x, t)\bar{q}(-x, t)q_x(x, t).
\eeq
\item The nonlocal DNLSIII equation:
\beq\label{nds3}
q_t(x, t)=\frac{i}{2}q_{xx}(x, t)\pm\frac{i}{2}q^2(x, t)\bar{q}_x(-x, t)-\frac{i}{4}q^3(x, t)\bar{q}^2(-x, t).
\eeq
\end{enumerate}

\begin{rem} Higher flows for each case can be carried out following the scheme in the previous sections. Hence we can get hierarchies of nonlocal DNLS-type equations.
\end{rem}

Hence we prove Theorem \ref{nn} stated in the very beginning of this section.

\begin{rem} We use the term ``nonlocal" to describe the similar constraints on DNLS equations as the case of NLS equation. It is an algebraic generalization instead of showing the dynamic properties as its original meaning in the NLS case.
\end{rem}

\section{Discussion}

In this paper, we generate hierarchies of DNLS-type (both local and nonlocal cases) from Lie algebra splittings. In particular, DNLSI, II, and III equations are derived from different loop algebra splittings. Moreover, we get a hierarchy of soliton equations for each DNLS-type. By choosing $G=U(1, 1)$, we derive new hierarchies of DNLS equations (\eqref{dds}, \eqref{dds2}, and \eqref{dds3}), which we call the defocusing analogies. Hence we give an algebraic explanation of the reductions of generalized KN hierarchies. In this manner, we prove that equations in the generalized KN hierarchies admit the constraints $r=\pm \bar{q}$. Under similar scheme, hierarchies of nonlocal DNLS-type equations (\eqref{nds1}, \eqref{nds2}, and \eqref{nds3}) are derived, and the nonlocal reductions ($r(x, t)=\pm i \bar{q}(-x, t)$) are induced by special automorphisms on the loop algebra of order four (\eqref{ae}, \eqref{ae2}, and \eqref{ae3}).

As mentioned in Sec. \ref{sp}, one advantage of the splitting theory is that, if a soliton hierarchy is constructed from Lie algebra splitting, then usually the B\"{a}cklund transformation on the space of solutions will follow from rational elements in the negative group $L_-$ with one or two poles and the process is algebraic. In a future work, we will concentrate on the B\"{a}cklund transformation for the DNLS-type equations introduced in this paper. We hope that we can get a clean scheme to obtain both soliton and rogue wave solutions for these equations. And we expect that we can generalize this process to nonlocal DNLS-type equations where we cannot apply the splitting theory directly.

Another interesting question to ask is whether all of these results can be generated to higher dimensions, i.e., to the vector DNLS-type equations. And in that case, the question is whether they admit similar nonlocal reductions. If so, another question is how to explain them in terms of algebra structure. Research in this direction is underway.

\begin{acknowledgement}
Zhiwei Wu is supported in part by NSF of China under Grant No. 11401327. Jingsong He
is supported in part by NSF of China under Grant No. 11271210 and the K.C. Wong Magna Fund in Ningbo University. Jingsong He thanks Prof. A.S. Fokas for arranging his visit to Cambridge University during 2012-2014 and for many useful discussions.
\end{acknowledgement}


\end{document}